\UseRawInputEncoding
\documentclass[12pt]{amsart}
\usepackage{amsmath}
\usepackage{amsthm}
\usepackage{amsfonts}
\usepackage{graphicx}
\usepackage{amssymb} 
\usepackage{bbold}

\newenvironment{nouppercase}{
  
  \renewcommand{\uppercasenonmath}[1]{}}{}

\begin{document}

\title[Classical dynamics of $SU(2)$ matrix models]
{Classical dynamics of $SU(2)$ matrix models}
\author[Jens Hoppe]{Jens Hoppe}
\address{Braunschweig University, Germany \& IHES, France}
\email{jens.r.hoppe@gmail.com}

\begin{abstract}
By direct, elementary, considerations it is shown that the $SU(2) \times SO(d=2,3)$ invariant sector of the bosonic membrane matrix model is governed by (2, resp. 3-dimensional) $x^2 y^2$-models.
\end{abstract}

\begin{nouppercase}
\maketitle
\end{nouppercase}
\thispagestyle{empty}
\noindent
While it has been known for more than 40 years \cite{1} that the relativistic membrane equations via a particular Ansatz reduce ￼ to the dynamics of a point particle experiencing a quartic $x^2y^2$ potential, and the effective potential for $SU(2) \times SO(d)$ invarant quantum mechanical eigenstates of the corresponding matrix-model(s) \cite{1} (resp. reduced Yang Mills theory \cite{2})  known equally long \cite{3}\cite{4}\footnote{
including an argument that for $d=3$ one of the invariant variables should be allowed to take negative values}\cite{5}\cite{6}\cite{7},
it is instructive to see a completely elementary derivation of the classical dynamics, whose
 $SU(2) \times SO(d=2,3)$ invariant motions are\footnote{at least 
 as long as the position vectors are linear independent}  governed by $x_i^2x_j^2$ potentials.
￼This ￼can be shown as follows: the equations of motion,
\begin{equation}\label{eq1} 
\begin{split}
\dot{\vec{p}}_1 & = \ddot{\vec{q}}_1 = (\vec{q}_1 \times \vec{q}_2) \times \vec{q}_2,\\
\dot{\vec{p}}_2 & = \ddot{\vec{q}}_2 = (\vec{q}_2 \times \vec{q}_1) \times \vec{q}_1,
\end{split}
\end{equation}
the $SU(2)$ gauge-constraint
\begin{equation}\label{eq2} 
\vec{J} := (\vec{q}_1 \times \vec{p}_1) + (\vec{q}_2 \times \vec{p}_2) = 0,
\end{equation}
and
\begin{equation}\label{eq3} 
(\dot{Q}=)P = (\vec{p}_1, \vec{p}_2) = (\vec{q}_1\vec{q}_2)\cdot A + (\vec{q}_1 \times \vec{q}_2)(\gamma_1, \gamma_2)
\end{equation}
(with A an arbitrary $2\times 2$ matrix; i.e. just using that $\vec{q}_1$, $\vec{q}_2$ and ($\vec{q}_1 \times \vec{q}_2$) can be used as a basis in $\mathbb{R}^3$, as long as $(\vec{q}_1 \times \vec{q}_2) \neq \vec{0}$) imply $\gamma_1 = \gamma_2 = 0$, $a_{12} = a_{21}$, i.e.
\begin{equation}\label{eq4} 
\dot{Q} = QA, \quad A^T = A;
\end{equation}
multiplying
\begin{equation}\label{eq5} 
Q(\dot{A} + A^2) = (QA)\dot{} = \ddot{Q} = Q 
\begin{pmatrix}
-\vec{q}^2_2 & \vec{q}_1\cdot\vec{q}_2 \\ \vec{q}_1\cdot\vec{q}_2 & -\vec{q}^2_1
\end{pmatrix}
= -Q|B|B^{-1} 
\end{equation}
by $(Q^T Q)^{-1} Q^T =: B^{-1}Q^T$ then gives
\begin{equation}\label{eq6} 
\dot{A} + A^2 + B^{-1} \text{det} B = 0,
\end{equation}
together with 
\begin{equation}\label{eq7} 
\dot{B} = (Q^T Q)\dot{} = Q^T \dot{Q} + \dot{Q}^T Q = BA +AB.
\end{equation}
This nice set of 2 first-order ODE’s for two symmetric $2 \times 2$ matrices (with $B$ positive definite , as by assumption $|B| = \text{det} B = (\vec{q}_1 \times \vec{q}_2)^2 \neq 0$) has an easy solution if, in addition to \eqref{eq2}, considering only $SO(2)$ invariant motions, i.e. assuming 
\begin{equation}\label{eq8}
\begin{split} 
0 \stackrel{!}{=} K & := \vec{q}_1 \vec{p}_2 - \vec{q}_2 \vec{p}_1 = (Q^T P - P^T Q)_{12}\\
& = (Q^T Q A - AQ^TQ)_{12} = [B, A]_{12},
\end{split}
\end{equation}
as then $[A,B] = K
\begin{pmatrix}
0 & -1 \\ 1 &0
\end{pmatrix}
 = 0$, so that $A$ and $B$ can be simultaneously diagonalized,
\begin{equation}\label{eq9}
\begin{split} 
B(t) & = R(t)\Lambda (t) R^{-1}(t), \\
A(t) & = R(t)D(t)R^{-1}(t).
\end{split}
\end{equation}
\eqref{eq7} then implies
\begin{equation}\label{eq10} 
\dot{\Lambda} - 2\Lambda D = [\Lambda, R^{-1}\dot{R}] \; (= 0)
\end{equation}
(where both sides have to be zero, as the left-hand side is diagonal, and the right-hand side purely off-diagonal; hence $\dot{R}$ or $\lambda_1-\lambda_2$ must vanish), while \eqref{eq6} gives 
\begin{equation}\label{eq11} 
\dot{d}_1 + d_1^2 + \lambda_2 = 0, \qquad \dot{d}_2 + d_2^2 + \lambda_1 = 0.
\end{equation}
$d_r=\frac{\dot{\psi}_r}{\psi_r}$ then gives
\begin{equation}\label{eq12} 
\ddot{\psi_1}+c_2^2 \psi_2^2 \psi_1=0  ,  \quad \ddot{\psi_2}+c_1^2 \psi_1^2 \psi_2=0 
\end{equation}
as $\dot{\Lambda}-2  \Lambda D = 0$ implies
\begin{equation}\label{eq13} 
\lambda_r=c_r^2 \psi_r^2 \quad (=\phi_r^2)
\end{equation}
and with
$\phi_r:= c_r \psi_r$ one gets
\begin{equation}\label{eq14} 
h(\phi_1 \phi_2 \pi_1 \pi_2):= \frac{1}{2}(\pi_1^2 + \pi_2^2 + \phi_1^2\phi_2^2)
\end{equation}
governing the dynamics ($\lambda_r$ being the square of $\phi_r$, and $d_r$ being its logarithmic derivative, $\frac{1}{2}(\ln \phi^2_r)\dot{}\,$).\\  
Perhaps one should note that (14) also arises as a result of a canonical point transformation motivated by the singular value decomposition,
\begin{equation}\label{eq15}
\begin{split} 
Q & =S(\theta_1,\theta_2,\theta_3) \begin{pmatrix}
\sigma_1 & 0 \\ 0 & \sigma_2 \\ 0& 0
\end{pmatrix} R(\theta) \\
& = (Q_{as})=Q(\phi^{\mu=1,2,...,6}=(\theta_1 \theta_2 \theta_3 \sigma_1 \sigma_2 \theta)) ,
\end{split}
\end{equation}
$Q,P \rightarrow \phi^{\mu} =\phi^{\mu}(Q), \Pi_{\mu} = \Pi_{\mu} (P,Q)$ ,
\begin{equation}\label{eq16} 
J=(J^{as}_{\mu} := \frac{\partial Q^{as}}{\partial \phi^{\mu}}), \, \Pi_{\mu}=J_{\mu}^{as}P_{as} = (J^T P)_{\mu} 
\end{equation}
with
\begin{equation}\label{eq17} 
\{\phi^{\mu}, \Pi_{\nu} \}=\delta_{\nu}^{\mu},\quad \{\Pi_{\mu}, \Pi_{\nu} \}=0 .
\end{equation}
$P = (J^T)^{-1}\pi$ then implies
\begin{equation}\label{eq18} 
H [\phi^{\mu}, \Pi_{\nu}]=\frac{1}{2}(\Pi_{\mu}G^{\mu \nu}(\phi)\Pi_{\nu}+\sigma_{1}^{2}\sigma_{2}^{2})
\end{equation}
where $(G^{\mu \nu})$ is the inverse of
\begin{equation}\label{eq19} 
(G_{\mu \nu}):=Tr(\partial_{\mu}Q^{T}\partial_{\nu}Q) ,
\end{equation}
with the (4,5) block being 
$\begin{pmatrix}
1&0 \\ 0&1
\end{pmatrix}$, and  $G_{4\,\text{or}\,5,\,a \,\text{or} \, 6}= 0$.\\
The $d=3$ case,
\begin{equation}\label{eq20} 
H=\frac{1}{2}(\vec{p}_1^2+\vec{p}_2^2+\vec{p}_3^2+(\vec{q}_{1}\times \vec{q}_2)^2+(\vec{q}_{2}\times \vec{q}_3)^2+(\vec{q}_{3}\times \vec{q}_1)^2) ,
\end{equation}
\begin{equation}\label{eq21} 
\ddot{\vec{q}}_i = \sum_{j=1}^{3}(\vec{q}_i \times \vec{q}_j)\times \vec{q}_j ,
\end{equation}
can be treated in more or less the same way; 
assuming $\vec{q}_1, \vec{q}_2, \vec{q}_3$ to be a basis (i.e. for $d=3$ taking $\vec{q}_3$,
rather than $\vec{q}_1 \times \vec{q}_2$), which not only excludes $V=0$, but also less degenerate configurations (namely those where the effective dimension is two, rather than three - or, when $V=0$, one), one may write
\begin{equation}\label{eq22} 
\dot{Q}= (\vec{p}_{1} \vec{p}_{2} \vec{p}_{3})= P = (\vec{q}_{1} \vec{q}_{2} \vec{q}_{3})A,
\end{equation}
which together with (21) implies
\begin{equation}\label{eq23} 
\dot{A}+A^2=B-\mathbb{1}(\text{Tr}B) ;
\end{equation}
also, again,
\begin{equation}\label{eq24} 
\dot{(Q^T Q)} =: \dot{B} =BA + AB .
\end{equation}
The $SU(2)$ constraints
\begin{equation}\label{eq25} 
\begin{aligned}
(\vec{J})_a & = (\sum_{i=1}^{d=3}\vec{q}_i \times \vec{p}_i)_a=(\epsilon_{abc}Q_{bi}Q_{cj})A_{ji}\\ 
& = -
\{(\vec{q}_1 \times \vec{q}_2)(A_{12}-A_{21})\\
& \quad +(\vec{q}_2 \times \vec{q}_3)(A_{23}-A_{32})+(\vec{q}_3 \times \vec{q}_1)(A_{31}-A_{13})\}_a=0
\end{aligned}
\end{equation}
imply that $A$ is symmetric, while the $SO(3)$ constraints,
\begin{equation}\label{eq26} 
0=K_{ij}= (\vec{q}_i \vec{p}_j-\vec{q}_j \vec{p}_i)=Q_{ai}P_{aj}-Q_{aj}P_{ai}=(B \cdot A)_{ij}-(A \cdot B)_{ij}
\end{equation}
then imply that $[A,B]$ (which, as the commutator of two real symmetric matrices is purely off-diagonal) has to vanish.
For the eigenvalues, $\sigma^2_i$,     
of $B$ being/assumed to be/different, the above not only implies that $A$ and $B$ can be simultaneously be diagonalized (which we now assume to be done) but also that the (orthogonal) ``diagonalisator" is time-independent, due to \eqref{eq23}/\eqref{eq24}. Hence,
with $A = \frac{\dot{\psi}}{\psi}$, $B=\Lambda$
(24) gives $\dot{\Lambda}\Psi = 2\Lambda\dot{\Psi}$, i.e.
\begin{equation}\label{eq27} 
\lambda_i=c_i^2 \Psi_i^2 ,
\end{equation}
and (23)
\begin{equation}\label{eq28} 
\ddot{\Psi}_i =c_i^2 \Psi_i^3-\Psi_i(\sum_j c_j^2 \Psi_j^2) ,
\end{equation}
i.e., for $\Phi_i := c_i\Psi_i$,
\begin{equation}\label{eq29} 
\ddot{\Phi}_1=-\Phi_1(\Phi_2^2+\Phi_3^2),  \ddot{\Phi}_2=-\Phi_2(\Phi_3^2+\Phi_1^2),  \ddot{\Phi}_3=-\Phi_3(\Phi_1^2+\Phi_2^2),
\end{equation}
which are Hamiltonian with respect to
\begin{equation}\label{eq30} 
h=\frac{1}{2}(\pi_1^2+\pi_2^2+\pi_3^2+\sum_{i<j}\Phi_i^2 \Phi_j^2) .
\end{equation}
It is important, however, to note that lower dimensional configurations do exist (that at first sight escape the above derivation), e.g.
\begin{equation}
\label{31}
\vec q_1 (t) = x(t) \vec e_1 \, , \ \vec q_2 (t) = \lambda x(t) \vec e_1 \, , \ \vec q_3 = z(t) \vec e_3
\end{equation}
$$
\ddot x = - x z^2 \, , \ \ddot z = -(1+\lambda^2) zx^2.
$$
These solutions are $SU(2) \times SO(3)$ invariant, but 
\begin{eqnarray}
\label{32}
B := Q^TQ &= &\begin{pmatrix}
x^2 &\lambda x^2 &0 \\
\lambda x^2 &\lambda^2 x^2 &0 \\
0 &0 &z^2
\end{pmatrix} \nonumber \\
&= &\begin{pmatrix}
\lambda &1 &0 \\
-1 &\lambda &0 \\
0 &0 &1
\end{pmatrix} \begin{pmatrix}
0 &0 &0 \\
0 &x^2 &0 \\
0 &0 &z^2
\end{pmatrix} \begin{pmatrix}
\lambda &-1 &0 \\
1 &\lambda &0 \\
0 &0 &1
\end{pmatrix}
\end{eqnarray}
has a vanishing eigenvalue (due to $\vec q_1 \times \vec q_2 \equiv 0$). Nevertheless
\begin{equation}
\label{33}
P = (\vec p_1 \vec p_2 \vec p_3) = (\vec q_1 \vec q_2 \vec q_3) \begin{pmatrix} 
\dot x / x & 0 & 0\\
0 &\dot x / x &0 \\
0 &0&\dot z/z
\end{pmatrix} = QA ,
\end{equation}
and
\begin{equation}
\label{34}
Q(\dot A + A^2) = Q (B - \mathbb{1}(\text{Tr}B)
\end{equation}
does hold (while (23) does {\it not}) despite of $\dot A + A^2 \ne B - \mathbb{1}(\text{Tr}B)$. Note that the upper $2 \times 2$ block of $Q$, in the natural basis $\vec e_1 , \vec e_2 , \vec e_3$, is $x \begin{pmatrix} 1 &\lambda \\ 0 &0 \end{pmatrix}$, and
\begin{equation}
\label{35}
\begin{pmatrix} c &s \\ -s &c \end{pmatrix} \begin{pmatrix} 1 &\lambda \\ 0 &0 \end{pmatrix} \begin{pmatrix} c &-s \\ s &c \end{pmatrix} = \begin{pmatrix} 0 &\lambda \\ 0 &1 \end{pmatrix}
\end{equation}
$$
c = \frac{\lambda}{\sqrt{1+\lambda^2}} \, , \ s = \frac{-1}{\sqrt{1+\lambda^2}} \, ,
$$
\begin{eqnarray}
\label{36}
&&\begin{pmatrix}
0 &\lambda x &0 \\
0 &x &0 \\
0 &0 &z
\end{pmatrix} \begin{pmatrix}
-z^2 - (1+x^2) x^2 & 0 &0 \\
0 &-z^2 &0 \\
0 & 0&-(1+\lambda^2)x^2
\end{pmatrix}  \\
&= &\begin{pmatrix}
0 &\lambda x &0 \\
0 &x &0 \\
0 &0 &z
\end{pmatrix} \begin{pmatrix}
\frac{\ddot x}x & 0 & 0 \\
&\frac{\ddot x}x &0 \\
0&0 &\frac{\ddot z}z
\end{pmatrix}. \nonumber
\end{eqnarray} 
(31) corresponds to the Lagrangian, resp. Hamiltonian
$${\mathcal L} = \frac12 \left[ (\dot z^2 + (1+ \lambda^2) \dot x^2) - (1+\lambda^2) x^2 z^2 \right]
$$
$${\mathcal H} = \frac12 \left[ (p^2 / (1+\lambda^2) + {\pi}^2 ) + (1+\lambda^2) x^2 z^2 \right] ;
$$ 
while the canonical transformation (simple rescaling) $\widetilde x = \sqrt{1+\lambda^2} \, x$, $\widetilde p = p/\sqrt{1+\lambda^2}$ reduces (31)/(37) to (14) (as if simply putting one pair of canonical variables in (30) equal to zero) it is important to note the physical significance of (31) (and the singularity in $A = \dot\Psi/\Psi$ if $\Psi_1 = 0$).

\vglue1cm

\noindent
\textbf{Acknowledgement}:
I would like to thank T.Anous, M.Bordemann, J.Fr\"ohlich, M.Hanada, and M.Hynek for discussions on related topics


\begin{thebibliography}{11111}
\bibitem[1]{1}  J.Hoppe, {\it Quantum Theory of a Massless Relativistic Surface}, MIT Ph.D. Thesis 1982,
http://dspace.mit.edu/handle/1721.1/15717
\bibitem[2]{2}S.G.Matinyan,G.K.Savvidy,N.G.Ter-Arutyunyan-Savvidy, {\it Classical Yang-Mills mechanics.Nonlinear color oscillations} Sov.Phys.JETP 53 1981
\bibitem[3]{3} J.Goldstone, J.Hoppe, unpublished notes 1980
\bibitem[4]{4} A.Smilga, {\it Witten Index Calculation in Supersymmetric Gauge Theory}, Nucl.Phys.B 266, 1986
\bibitem[5]{5} M.Halpern, C.Schwarz, {\it Asymptotic Search for Ground States of SU(2) Matrix Theory} Int.J.Mod.Phys.A13, 1998
\bibitem[6]{6} J.Hoppe, {\it On Zero-Mass Bound States in Supermembrane-Models} arXiv:hep-th/9609232 (1996),
{\it Zero energy states in supersymmetric matrix models}, Class.Quant.Grav.17 (2000) 
\bibitem[7]{7} T.Anous, C.Cogburn, {\it Mini-BFSS matrix model in silico}, Phys.Rev.D 100 (2019)
\end{thebibliography}
\end{document}